\documentclass[prd,nopacs,preprintnumbers,twocolumn,amsmath,nofootinbib,amssymb]{revtex4}
\usepackage{graphicx,color,dcolumn,booktabs,bm}
\usepackage{longtable,lscape}
\usepackage{txfonts}
\usepackage{overpic}
\usepackage{epstopdf}
\usepackage{indentfirst}
\usepackage{threeparttable}
\usepackage{feynmf}   
\usepackage{slashed}  
\usepackage{cases}
\usepackage{color}
\usepackage{float}
\usepackage{makecell}
\usepackage{multirow}
\usepackage{array}
\usepackage{gensymb}
\usepackage{graphicx,color,dcolumn,booktabs,bm}
\usepackage{epsfig,dsfont,amssymb,amsmath,amsfonts,amsbsy,mathrsfs}

\graphicspath{{Figures/}} %

\usepackage{hyperref}
\hypersetup{colorlinks,citecolor=blue,anchorcolor=red,menucolor=red, linkcolor=red,filecolor=red,runcolor=red,urlcolor=blue,frenchlinks=red}

\makeatletter
\@addtoreset{equation}{section}
\makeatother

\allowdisplaybreaks

\newcolumntype{I}{!{\vrule width 0.9pt}}

\begin{document}

\title{Revisiting heavy-flavor conserving weak decays of charmed baryons within the nonrelativistic quark model}
\author{Yu-Shuai Li$^{1}$}\email{liysh@pku.edu.cn}
\author{Xiang Liu$^{2,3,4}$}\email{xiangliu@lzu.edu.cn}
\affiliation{$^1$ School of Physics and Center of High Energy Physics, Peking University, Beijing 100871, China\\
$^2$Lanzhou Center for Theoretical Physics,
Key Laboratory of Theoretical Physics of Gansu Province,
Key Laboratory of Quantum Theory and Applications of MoE,
Gansu Provincial Research Center for Basic Disciplines of Quantum Physics, Lanzhou University, Lanzhou 730000, China\\
$^3$MoE Frontiers Science Center for Rare Isotopes, Lanzhou University, Lanzhou 730000, China\\
$^4$Research Center for Hadron and CSR Physics, Lanzhou University and Institute of Modern Physics of CAS, Lanzhou 730000, China}

\begin{abstract}

In this work, we investigate the heavy-flavor conserving weak decays of charmed baryons, specifically the processes $\Xi_{c} \to \Lambda_{c} \pi$ and $\Omega_{c} \to \Xi_{c} \pi$, where the pole model is employed to account for the nonfactorizable contributions. Additionally, the nonperturbative parameters involved are determined within the framework of the nonrelativistic quark model, utilizing exact baryon wave functions obtained by solving the Schr\"{o}dinger equation with a nonrelativistic potential. By considering the mixing angle $\theta \in (24.4^\circ, 32.4^\circ)$ for $\Xi_{c}-\Xi_{c}^{\prime}$ mixing, we find that the experimental data can be effectively reproduced. Furthermore, we estimate the branching ratios: $\mathcal{B}(\Xi_{c}^{+} \to \Lambda_{c}^{+} \pi^{0}) = (8.69 \sim 9.79) \times 10^{-3}$, $\mathcal{B}(\Omega_{c}^{0} \to \Xi_{c}^{+} \pi^{-}) = (11.3 \sim 12.1) \times 10^{-3}$, and $\mathcal{B}(\Omega_{c}^{0} \to \Xi_{c}^{0} \pi^{0}) = (4.67 \sim 5.23) \times 10^{-3}$. These predictions can be tested in ongoing experiments at LHCb, Belle II, and BESIII.
\end{abstract}

\maketitle

\section{introduction}
\label{sec1}

\begin{figure}[htbp]\centering
  \includegraphics[width=82mm]{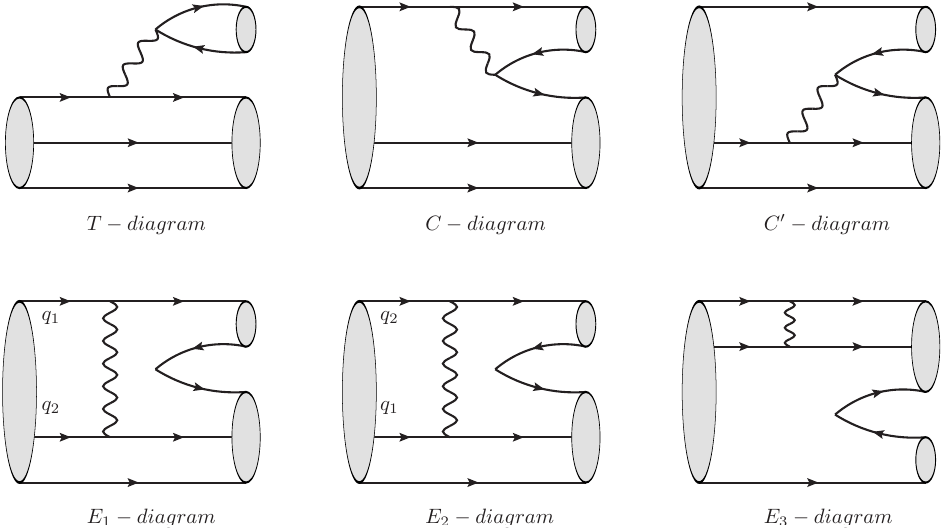}\\
  \caption{The topological diagrams for baryon weak decays include the factorizable external $W$-emission diagram ($T$) and internal $W$-emission diagram ($C$), as well as the nonfactorizable inner $W$-emission diagram ($C^{\prime}$) and $W$-exchange diagrams ($E_{1,2,3}$).}
  \label{fig:Topological Diagrams}
\end{figure}

The study of nonleptonic decays of charmed baryons offers valuable insights into both the weak decay mechanisms and the nonperturbative effects in QCD. However, due to the complexity of the nonperturbative nature of low-energy QCD, theoretical developments have been challenging and remain incomplete. For charmed baryon nonleptonic decays, as illustrated in Fig.~\ref{fig:Topological Diagrams}, the topological diagrams consist of the external $W$-emission diagram ($T$), the internal $W$-emission diagram ($C$), the inner $W$-emission diagram ($C^{\prime}$), and the $W$-exchange diagrams ($E_{1,2,3}$), at tree level. Among these, $T$ and $C$ are factorizable, while $C^{\prime}$ and $E_{1,2,3}$ are nonfactorizable \cite{Cheng:2021qpd}. The factorizable contributions can often be estimated under the na\"{i}ve factorization assumption, though this is a rough approximation. In contrast, the nonfactorizable contributions are more difficult to calculate. To address this, several approaches have been developed, including the pole model~\cite{Cheng:1991sn, Cheng:1992ff, Cheng:1993gf, Uppal:1994pt, Cheng:2018hwl, Zou:2019kzq, Hu:2020nkg, Cheng:2020wmk, Liu:2022igi, Meng:2020euv, Niu:2021qcc, Cheng:2022kea, Zeng:2022egh, Cheng:2022jbr, Ivanov:2023wir}, QCD sum rules (QCDSR)~\cite{Shi:2022kfa, Shi:2024plf}, and SU(3) flavor symmetry~\cite{Sharma:1996sc, Lu:2016ogy, Geng:2017esc, Geng:2018plk, Geng:2018rse, Geng:2019xbo, Geng:2019awr, Groote:2021pxt, Zhong:2022exp, Xing:2023dni, Geng:2024sgq, Wang:2024ztg}, which are employed to study these decays.

Particularly for charmed baryon nonleptonic decays, the heavy-flavor-conserving (HFC) processes $\Xi_{c} \to \Lambda_c \pi$ and $\Omega_{c} \to \Xi_c \pi$ provide a simpler way to investigate the underlying mechanisms. To assess the nonfactorizable contributions, a series of theoretical studies have been conducted using the pole model~\cite{Cheng:1992ff, Niu:2021qcc, Cheng:2022kea, Cheng:2022jbr, Ivanov:2023wir}. In the framework of the pole model, the nonfactorizable S-wave parity-violating (PV) amplitudes primarily arise from low-lying $J^{P}=1/2^{-}$ poles, while nonfactorizable P-wave parity-conserving (PC) amplitudes are dominated by ground-state poles~\cite{Cheng:1991sn}. Within this approach, the authors of Ref.~\cite{Niu:2021qcc} evaluated the contributions to nonfactorizable PV amplitudes by considering various negative-parity charmed baryons, using the nonrelativistic quark model (NRQM). However, for HFC weak decays, the application of the soft-pion theorem—guaranteed by the small mass gaps between the charmed baryons—leads to a simplification in the current algebra approach, bypassing the negative-parity $J^{P}=1/2^{-}$ poles and significantly reducing the difficulty of calculation~\cite{Cheng:2022kea, Cheng:2022jbr}.

The pole model has been successfully applied to evaluate the nonfactorizable contributions in the nonleptonic decays of charmed baryons~\cite{Cheng:1991sn, Cheng:1992ff, Cheng:1993gf, Uppal:1994pt, Cheng:2018hwl, Zou:2019kzq, Hu:2020nkg, Cheng:2020wmk, Liu:2022igi, Meng:2020euv, Niu:2021qcc, Cheng:2022kea, Zeng:2022egh, Cheng:2022jbr, Ivanov:2023wir}. For calculating the nonperturbative parameters in decay amplitudes, various phenomenological quark models have been employed, including the bag model~\cite{Cheng:1991sn, Cheng:1992ff, Cheng:1993gf, Cheng:2018hwl, Zou:2019kzq, Hu:2020nkg, Cheng:2020wmk, Liu:2022igi, Meng:2020euv, Cheng:2022kea, Cheng:2022jbr}, the diquark model~\cite{Cheng:1992ff, Cheng:2022kea}, NRQM~\cite{Niu:2021qcc, Zeng:2022egh}, and the covariant confined quark model (CCQM)~\cite{Ivanov:2023wir}.

In this work, we revisit the HFC weak decays $\Xi_{c} \to \Lambda_c \pi$ and $\Omega_{c} \to \Xi_c \pi$ using the pole model and evaluate the relevant nonperturbative parameters within the NRQM framework. In our numerical calculations, the primary uncertainties stem from the input nonperturbative parameters, especially the baryon-baryon matrix element. To address this challenge, we propose using exact baryon spatial wave functions, obtained by solving the Schr\"{o}dinger equation with a nonrelativistic potential, rather than relying on an oversimplified Gaussian-type approximation. This approach constitutes the main contribution of our study. With support from charmed baryon spectroscopy, our strategy minimizes the dependence on arbitrary wave functions for nonperturbative parameters, thereby reducing the associated uncertainties.

The paper is organized as follows: After the Introduction, we calculate the decay amplitudes in Sec.~\ref{sec2}, where the nonfactorizable amplitudes are evaluated using the pole model and simplified via the soft-pion approximation. The nonperturbative parameters are determined within the NRQM framework. In Sec.~\ref{sec3}, we introduce a nonrelativistic potential and solve the Schr\"{o}dinger equation to obtain the exact baryon wave functions, which are then used to calculate the relevant nonperturbative parameters. In Sec.~\ref{sec4}, we present the numerical results for the nonperturbative parameters and investigate the physical observables for the weak decays under study. Finally, Sec.~\ref{sec5} provides a brief summary of the work.

\section{The HFC weak decay amplitudes of charmed baryons}
\label{sec2}

\subsection{The factorizable and nonfactorizable amplitudes in pole model}
\label{sec2.1}

The effective Hamiltonians relevant to the HFC weak decays under consideration are given by
\begin{equation}
\mathcal{H}_{1,\text{eff}}=\frac{G_{F}}{\sqrt{2}}V_{us}V_{ud}^{*}\Big{(}c_{1}O_{1}+c_{2}O_{2}\Big{)}+\text{H.c.}
\end{equation}
with $O_{1}=(\bar{d}u)(\bar{u}s)$ and $O_{2}=(\bar{u}u)(\bar{d}s)$, and
\begin{equation}
\mathcal{H}_{2,\text{eff}}=\frac{G_{F}}{\sqrt{2}}V_{cs}V_{cd}^{*}\Big{(}c_{1}\tilde{O}_{1}+c_{2}\tilde{O}_{2}\Big{)}+\text{H.c.}
\end{equation}
with $\tilde{O}_{1}=(\bar{d}c)(\bar{c}s)$ and $\tilde{O}_{2}=(\bar{c}c)(\bar{d}s)$. Here, the four-fermion operator are denoted as $(\bar{q}{1}q{2}) = \bar{q}{1} \gamma{\mu}(1 - \gamma_{5}) q_{2}$. In this work, we adopt the effective Wilson coefficients $c_{1} = 1.336$ and $c_{2} = -0.621$~\cite{Cheng:2022kea}. The Cabibbo-Kobayashi-Maskawa (CKM) matrix elements are expressed as
\begin{equation*}
\begin{split}
V_{ud}=&1-\lambda^{2}/2,V_{us}=\lambda,\\
V_{cd}=&-\lambda,V_{cs}=1-\lambda^{2}/2,
\end{split}
\end{equation*}
where $\lambda = 0.22501$~\cite{ParticleDataGroup:2024cfk}. It is important to emphasize that $\mathcal{H}_{2, \text{eff}}$ corresponds to the quark-level transition $cs \to dc$, leading to non-negligible nonspectator $W$-exchange contributions~\cite{Gronau:2016xiq, Niu:2021qcc, Cheng:2022kea, Cheng:2022jbr}.

For describing the two-body nonleptonic decay $\mathcal{B}{i} \to \mathcal{B}{f} + P~(P \equiv \text{pseudoscalar meson})$, the total amplitude can be parameterized as the sum of two terms:
\begin{equation}
\mathcal{M}(\mathcal{B}_{i}\to\mathcal{B}_{f}+P)=i\bar{u}_{f}(A-B\gamma_{5})u_{i},
\end{equation}
where $A$ and $B$ represent the PV and PC amplitudes, respectively. Each amplitude receives both factorizable and nonfactorizable contributions, namely:
\begin{equation}
A=A^{\text{fac}}+A^{\text{nf}},~B=B^{\text{fac}}+B^{\text{nf}}.
\end{equation}

Under the na\"{i}ve factorization assumption, the factorizable amplitude can be easily evaluated by factorizing the hadronic matrix element into the product of two separate matrix elements. In the study of heavy-flavor-conserving decays $\Xi_{c}^{0,+}\to\Lambda_{c}\pi^{-,0}$, the $\Xi_{c} - \Xi_{c}^{\prime}$ mixing, described by
\begin{equation}
\begin{split}
\vert\Xi_{c}\rangle=&\cos\theta\vert\Xi_{c}^{\bar{3}}\rangle+\sin\theta\vert\Xi_{c}^{6}\rangle,\\
\vert\Xi_{c}^{\prime}\rangle=&-\sin\theta\vert\Xi_{c}^{\bar{3}}\rangle+\cos\theta\vert\Xi_{c}^{6}\rangle
\end{split}
\end{equation}
where $\Xi_{c}$ and $\Xi_{c}^{\prime}$ represent the physical states, and $\Xi_{c}^{\bar{3}}$ and $\Xi_{c}^{6}$ denote the antitriplet and sextet states, respectively, plays a significant role in the P-wave amplitudes. This mechanism has been proposed in Ref.~\cite{Cheng:1992ff}. After considering the $\Xi_{c} - \Xi_{c}^{\prime}$ mixing, the factorizable amplitudes, within the na\"{i}ve factorization assumption, can be expressed as:
\begin{widetext}
\begin{equation}
\begin{split}
A^{\text{fac}}(\Xi_{c}^{0}\to\Lambda_{c}^{+}\pi^{-})=&-\frac{G_{F}}{\sqrt{2}}V_{us}V_{ud}^{*}a_{1}f_{\pi}
\bigg{[}f_{1}^{\Lambda_c^{+}\Xi_c^{\bar{3},0}}(m_{\pi}^{2})\cos\theta+f_{1}^{\Lambda_c^{+}\Xi_c^{6,0}}(m_{\pi}^{2})\sin\theta\bigg{]}(m_{i}-m_{f}),\\
B^{\text{fac}}(\Xi_{c}^{0}\to\Lambda_{c}^{+}\pi^{-})=&\frac{G_{F}}{\sqrt{2}}V_{us}V_{ud}^{*}a_{1}f_{\pi}
\bigg{[}g_{1}^{\Lambda_c^{+}\Xi_c^{\bar{3},0}}(m_{\pi}^{2})\cos\theta+g_{1}^{\Lambda_c^{+}\Xi_c^{6,0}}(m_{\pi}^{2})\sin\theta\bigg{]}(m_{i}+m_{f}),
\end{split}
\end{equation}
\begin{equation}
\begin{split}
A^{\text{fac}}(\Xi_{c}^{+}\to\Lambda_{c}^{+}\pi^{0})=&-\frac{G_{F}}{2}V_{us}V_{ud}^{*}a_{2}f_{\pi}
\bigg{[}f_{1}^{\Lambda_c^{+}\Xi_c^{\bar{3},+}}(m_{\pi}^{2})\cos\theta+f_{1}^{\Lambda_c^{+}\Xi_c^{6,+}}(m_{\pi}^{2})\sin\theta\bigg{]}(m_{i}-m_{f}),\\
B^{\text{fac}}(\Xi_{c}^{+}\to\Lambda_{c}^{+}\pi^{0})=&\frac{G_{F}}{2}V_{us}V_{ud}^{*}a_{2}f_{\pi}
\bigg{[}g_{1}^{\Lambda_c^{+}\Xi_c^{\bar{3},+}}(m_{\pi}^{2})\cos\theta+g_{1}^{\Lambda_c^{+}\Xi_c^{6,+}}(m_{\pi}^{2})\sin\theta\bigg{]}(m_{i}+m_{f}),
\end{split}
\end{equation}
\begin{equation}
\begin{split}
A^{\text{fac}}(\Omega_{c}^{0}\to\Xi_{c}^{+}\pi^{-})=&-\frac{G_{F}}{\sqrt{2}}V_{us}V_{ud}^{*}a_{1}f_{\pi}
\bigg{[}f_{1}^{\Xi_c^{\bar{3},+}\Omega_c^{0}}(m_{\pi}^{2})\cos\theta+f_{1}^{\Xi_c^{6,+}\Omega_c^{0}}(m_{\pi}^{2})\sin\theta\bigg{]}(m_{i}-m_{f}),\\
B^{\text{fac}}(\Omega_{c}^{0}\to\Xi_{c}^{+}\pi^{-})=&\frac{G_{F}}{\sqrt{2}}V_{us}V_{ud}^{*}a_{1}f_{\pi}
\bigg{[}g_{1}^{\Xi_c^{\bar{3},+}\Omega_c^{0}}(m_{\pi}^{2})\cos\theta+g_{1}^{\Xi_c^{6,+}\Omega_c^{0}}(m_{\pi}^{2})\sin\theta\bigg{]}(m_{i}+m_{f}),
\end{split}
\end{equation}
\begin{equation}
\begin{split}
A^{\text{fac}}(\Omega_{c}^{0}\to\Xi_{c}^{0}\pi^{0})=&-\frac{G_{F}}{2}V_{us}V_{ud}^{*}a_{2}f_{\pi}
\bigg{[}f_{1}^{\Xi_c^{\bar{3},0}\Omega_c^{0}}(m_{\pi}^{2})\cos\theta+f_{1}^{\Xi_c^{6,0}\Omega_c^{0}}(m_{\pi}^{2})\sin\theta\bigg{]}(m_{i}-m_{f}),\\
B^{\text{fac}}(\Omega_{c}^{0}\to\Xi_{c}^{0}\pi^{0})=&\frac{G_{F}}{2}V_{us}V_{ud}^{*}a_{2}f_{\pi}
\bigg{[}g_{1}^{\Xi_c^{\bar{3},0}\Omega_c^{0}}(m_{\pi}^{2})\cos\theta+g_{1}^{\Xi_c^{6,0}\Omega_c^{0}}(m_{\pi}^{2})\sin\theta\bigg{]}(m_{i}+m_{f}),
\end{split}
\end{equation}
\end{widetext}
for $\Xi_{c}^{0}\to\Lambda_{c}^{+}\pi^{-}$, $\Xi_{c}^{+}\to\Lambda_{c}^{+}\pi^{0}$, $\Omega_{c}^{0}\to\Xi_{c}^{+}\pi^{-}$ and $\Omega_{c}^{0}\to\Xi_{c}^{0}\pi^{0}$ weak decays, respectively, where $m_{i}~(m_{f})$ is the mass of initial (final) baryon, and the effective Wilson coefficients are given by $a_{1} = c_{1} + \frac{c_{2}}{N_{\text{eff}}}$ and $a_{2} = c_{2} + \frac{c_{1}}{N_{\text{eff}}}$, with $N_{\text{eff}} \approx 7$ determined from the $\Lambda_{c} \to p\phi$ process~\cite{Cheng:2018hwl}. The baryon transition form factors are defined by
\begin{equation}
\begin{split}
\langle\mathcal{B}_{f}&(P^{\prime})\vert\bar{q}^{\prime}\gamma_{\mu}(1-\gamma_{5})q\vert\mathcal{B}_{i}(P)\rangle=\\
&\bar{u}_{f}\bigg{\{}f_{1}^{\mathcal{B}_{f}\mathcal{B}_{i}}(q^{2})\gamma_{\mu}
+f_{2}^{\mathcal{B}_{f}\mathcal{B}_{i}}(q^{2})\frac{i\sigma_{\mu\nu}q^{\nu}}{m_{i}}
+f_{3}^{\mathcal{B}_{f}\mathcal{B}_{i}}(q^{2})\frac{q_{\mu}}{m_{i}}\\
&-\Big{[}g_{1}^{\mathcal{B}_{f}\mathcal{B}_{i}}(q^{2})\gamma_{\mu}
+g_{2}^{\mathcal{B}_{f}\mathcal{B}_{i}}(q^{2})\frac{i\sigma_{\mu\nu}q^{\nu}}{m_{i}}
+g_{3}^{\mathcal{B}_{f}\mathcal{B}_{i}}(q^{2})\frac{q_{\mu}}{m_{i}}\Big{]}\gamma_{5}\bigg{\}}u_{i},
\end{split}
\end{equation}
and the decay constant of the pseudoscalar meson is defined by
\begin{equation}
\langle{P(q)}\vert\bar{q}^{\prime}\gamma_{\mu}\gamma_{5}q\vert0\rangle=if_{P}q_{\mu}
\end{equation}
with $q\equiv P-P^{\prime}$. The form factors $f_{3}$ and $g_{3}$ are neglected in this calculation.

The pole model is employed to handle the nonfactorizable contributions arising from the inner $W$-emission diagram ($C^{\prime}$) and the $W$-exchange diagrams ($E_{1}$, $E_{2}$, and $E_{3}$). Based on Refs.~\cite{Cheng:1993gf, Cheng:2018hwl, Cheng:2021qpd}, the nonfactorizable PV and PC amplitudes are expressed as
\begin{equation}
\begin{split}
A^{\text{nf}}=&-\sum_{\mathcal{B}_{n^{*}}}\Bigg{(}\frac{g_{\mathcal{B}_{f}\mathcal{B}_{n^{*}}P}b_{n^{*}i}}{m_{i}-m_{n^{*}}}+\frac{b_{fn^{*}}g_{\mathcal{B}_{n^{*}}\mathcal{B}_{i}P}}{m_{f}-m_{n^{*}}}\Bigg{)}+\cdots,\\
B^{\text{nf}}=&-\sum_{\mathcal{B}_{n}}\Bigg{(}\frac{g_{\mathcal{B}_{f}\mathcal{B}_{i}P}a_{ni}}{m_{i}-m_{n}}+\frac{a_{fn}g_{\mathcal{B}_{n}\mathcal{B}_{i}P}}{m_{f}-m_{n}}\Bigg{)}+\cdots,
\end{split}
\end{equation}
where the amplitudes $a_{fi}$ and $b_{fi}$ are defined by the baryon-baryon matrix element:
\begin{equation}
\begin{split}
\langle\mathcal{B}_{f}\vert\mathcal{H}_{\text{eff}}\vert\mathcal{B}_{i}\rangle=&\bar{u}_{f}(a_{fi}+b_{fi}\gamma_{5})u_{i},\\
\langle\mathcal{B}_{f^{*}}\vert\mathcal{H}_{\text{eff}}^{\text{PV}}\vert\mathcal{B}_{i}\rangle=&b_{f^{*}i}\bar{u}_{f^{*}}u_{i}.
\label{eq:baryonmatrixelement}
\end{split}
\end{equation}

In HFC weak decays, the small mass difference between the initial and final charmed baryons ensures that the emitted pseudoscalar meson is soft~\cite{Cheng:2022kea, Cheng:2022jbr}. Under the soft-pion approximation, the nonfactorizable PV amplitudes can be evaluated as
\begin{widetext}
\begin{eqnarray}
A^{\text{nf}}(\Xi_{c}^{0}\to\Lambda_{c}^{+}\pi^{-})&=&\frac{1}{f_{\pi}}(a_{\Lambda_{c}^{+}\Xi_{c}^{\bar{3},+}}\cos\theta+\tilde{a}_{\Lambda_{c}^{+}\Xi_{c}^{\bar{3},+}}\cos\theta+\tilde{a}_{\Lambda_{c}^{+}\Xi_{c}^{6,+}}\sin\theta),\\
A^{\text{nf}}(\Xi_{c}^{+}\to\Lambda_{c}^{+}\pi^{0})&=&\frac{1}{\sqrt{2}f_{\pi}}(a_{\Lambda_{c}^{+}\Xi_{c}^{\bar{3},+}}\cos\theta+\tilde{a}_{\Lambda_{c}^{+}\Xi_{c}^{\bar{3},+}}\cos\theta+\tilde{a}_{\Lambda_{c}^{+}\Xi_{c}^{6,+}}\sin\theta),\\
A^{\text{nf}}(\Omega_{c}^{0}\to\Xi_{c}^{+}\pi^{-})&=&-\frac{1}{f_{\pi}}(\tilde{a}_{\Xi_{c}^{\bar{3},0}\Omega_{c}^{0}}\cos\theta+\tilde{a}_{\Xi_{c}^{6,0}\Omega_{c}^{0}}\sin\theta),\\
A^{\text{nf}}(\Omega_{c}^{0}\to\Xi_{c}^{0}\pi^{0})&=&\frac{1}{\sqrt{2}f_{\pi}}(\tilde{a}_{\Xi_{c}^{\bar{3},0}\Omega_{c}^{0}}\cos\theta+\tilde{a}_{\Xi_{c}^{6,0}\Omega_{c}^{0}}\sin\theta),
\end{eqnarray}
if considering the intermediate charmed baryons $\mathcal{B}_{n}=(\Lambda_{c}^{+},\Sigma_{c}^{0},\Sigma_{c}^{+},\Sigma_{c}^{++},\Xi_{c}^{\bar{3},0},\Xi_{c}^{\bar{3},+},\Xi_{c}^{6,0},\Xi_{c}^{6,+},\Omega_{c}^{0})$. The nonzero contributions to the nonfactorizable PC amplitudes are
\begin{equation}
\begin{split}
B^{\text{nf}}(\Xi_{c}^{0}\to\Lambda_{c}^{+}\pi^{-})=&-\frac{1}{f_{\pi}}\Bigg{[}
g_{\Lambda_{c}^{+}\Sigma_{c}^{0}}^{A(\pi^{-})}\frac{m_{\Lambda_{c}^{+}}+m_{\Sigma_{c}^{0}}}{m_{\Xi_{c}^{0}}-m_{\Sigma_{c}^{0}}}
(\tilde{a}_{\Sigma_{c}^{0}\Xi_{c}^{\bar{3},0}}\cos\theta+\tilde{a}_{\Sigma_{c}^{0}\Xi_{c}^{6,0}}\sin\theta)\\
&+(a_{\Lambda_{c}^{+}\Xi_{c}^{\bar{3},+}}+\tilde{a}_{\Lambda_{c}^{+}\Xi_{c}^{\bar{3},+}})\frac{m_{\Xi_{c}^{0}}+m_{\Xi_{c}^{\bar{3},+}}}{m_{\Lambda_{c}^{+}}-m_{\Xi_{c}^{\bar{3},+}}}
g_{\Xi_{c}^{\bar{3},+}\Xi_{c}^{6,0}}^{A(\pi^{-})}\sin\theta
+\tilde{a}_{\Lambda_{c}^{+}\Xi_{c}^{6,+}}\frac{m_{\Xi_{c}^{0}}+m_{\Xi_{c}^{6,+}}}{m_{\Lambda_{c}^{+}}-m_{\Xi_{c}^{6,+}}}
(g_{\Xi_{c}^{6,+}\Xi_{c}^{\bar{3},0}}^{A(\pi^{-})}\cos\theta+g_{\Xi_{c}^{6,+}\Xi_{c}^{6,0}}^{A(\pi^{-})}\sin\theta)
\Bigg{]},
\label{eq:Bnf1}
\end{split}
\end{equation}
\begin{equation}
\begin{split}
B^{\text{nf}}(\Xi_{c}^{+}\to\Lambda_{c}^{+}\pi^{0})=&-\frac{\sqrt{2}}{f_{\pi}}\Bigg{[}
g_{\Lambda_{c}^{+}\Sigma_{c}^{+}}^{A(\pi^{0})}\frac{m_{\Lambda_{c}^{+}}+m_{\Sigma_{c}^{+}}}{m_{\Xi_{c}^{+}}-m_{\Sigma_{c}^{+}}}
(\tilde{a}_{\Sigma_{c}^{+}\Xi_{c}^{\bar{3},+}}\cos\theta+\tilde{a}_{\Sigma_{c}^{+}\Xi_{c}^{6,+}}\sin\theta)\\
&+(a_{\Lambda_{c}^{+}\Xi_{c}^{\bar{3},+}}+\tilde{a}_{\Lambda_{c}^{+}\Xi_{c}^{\bar{3},+}})\frac{m_{\Xi_{c}^{+}}+m_{\Xi_{c}^{\bar{3},+}}}{m_{\Lambda_{c}^{+}}-m_{\Xi_{c}^{\bar{3},+}}}
g_{\Xi_{c}^{\bar{3},+}\Xi_{c}^{6,+}}^{A(\pi^{0})}\sin\theta
+\tilde{a}_{\Lambda_{c}^{+}\Xi_{c}^{6,+}}\frac{m_{\Xi_{c}^{+}}+m_{\Xi_{c}^{6,+}}}{m_{\Lambda_{c}^{+}}-m_{\Xi_{c}^{6,+}}}
(g_{\Xi_{c}^{6,+}\Xi_{c}^{\bar{3},+}}^{A(\pi^{0})}\cos\theta+g_{\Xi_{c}^{6,+}\Xi_{c}^{6,+}}^{A(\pi^{0})}\sin\theta)
\Bigg{]},
\label{eq:Bnf2}
\end{split}
\end{equation}
\begin{equation}
\begin{split}
B^{\text{nf}}(\Omega_{c}^{0}\to\Xi_{c}^{+}\pi^{-})=-\frac{1}{f_{\pi}}\Bigg{[}
(g_{\Xi_{c}^{\bar{3},+}\Xi_{c}^{6,0}}^{A(\pi^{-})}\cos\theta+g_{\Xi_{c}^{6,+}\Xi_{c}^{6,0}}^{A(\pi^{-})}\sin\theta)\frac{m_{\Xi_{c}^{+}}+m_{\Xi_{c}^{6,0}}}{m_{\Omega_{c}^{0}}-m_{\Xi_{c}^{6,0}}}\tilde{a}_{\Xi_{c}^{6,0}\Omega_{c}^{0}}
+g_{\Xi_{c}^{6,+}\Xi_{c}^{\bar{3},0}}^{A(\pi^{-})}\sin\theta\frac{m_{\Xi_{c}^{+}}+m_{\Xi_{c}^{\bar{3},0}}}{m_{\Omega_{c}^{0}}-m_{\Xi_{c}^{\bar{3},0}}}\tilde{a}_{\Xi_{c}^{\bar{3},0}\Omega_{c}^{0}}
\Bigg{]},
\label{eq:Bnf3}
\end{split}
\end{equation}
\begin{equation}
\begin{split}
B^{\text{nf}}(\Omega_{c}^{0}\to\Xi_{c}^{0}\pi^{0})=-\frac{\sqrt{2}}{f_{\pi}}\Bigg{[}
(g_{\Xi_{c}^{\bar{3},0}\Xi_{c}^{6,0}}^{A(\pi^{0})}\cos\theta+g_{\Xi_{c}^{6,0}\Xi_{c}^{6,0}}^{A(\pi^{0})}\sin\theta)\frac{m_{\Xi_{c}^{0}}+m_{\Xi_{c}^{6,0}}}{m_{\Omega_{c}^{0}}-m_{\Xi_{c}^{6,0}}}\tilde{a}_{\Xi_{c}^{6,0}\Omega_{c}^{0}}
+g_{\Xi_{c}^{6,0}\Xi_{c}^{\bar{3},0}}^{A(\pi^{0})}\sin\theta\frac{m_{\Xi_{c}^{0}}+m_{\Xi_{c}^{\bar{3},0}}}{m_{\Omega_{c}^{0}}-m_{\Xi_{c}^{\bar{3},0}}}\tilde{a}_{\Xi_{c}^{\bar{3},0}\Omega_{c}^{0}}
\Bigg{]}.
\label{eq:Bnf4}
\end{split}
\end{equation}
\end{widetext}
Here, $f_{\pi} = 130$ MeV is the decay constant of the pion, and the baryonic matrix elements $a_{\mathcal{B}^{\prime}\mathcal{B}}$ and $\tilde{a}_{\mathcal{B}^{\prime}\mathcal{B}}$ are defined by Eq.~\eqref{eq:baryonmatrixelement}, with $\mathcal{H}$ corresponding to $\mathcal{H}_{1,\text{eff}}$ and $\mathcal{H}_{2,\text{eff}}$, respectively. It is worthy to mention that, the baryonic matrix elements $a_{\mathcal{B}\mathcal{B}^{\prime}}$ for antitriplet-sextet charmed baryon transitions are vanish due to the heavy quark symmetry~\cite{Cheng:1992ff,Cheng:2022kea}.

\subsection{The nonperturbative parameters in NRQM}
\label{sec2.2}

Clearly, three types of nonperturbative parameters—namely, the baryon transition form factors $f_{1}^{\mathcal{B}^{\prime}\mathcal{B}}$ and $g_{1}^{\mathcal{B}^{\prime}\mathcal{B}}$, the axial-vector form factor $g_{\mathcal{B}^{\prime}\mathcal{B}}^{A}$, and the baryonic matrix elements $a_{\mathcal{B}^{\prime}\mathcal{B}}$ and $\tilde{a}_{\mathcal{B}^{\prime}\mathcal{B}}$—need to be determined. In this work, we evaluate these nonperturbative parameters within the framework of NRQM.

Typically, a baryon state in momentum space can be expressed in terms of mock states as
\begin{equation}
\begin{split}
\vert\mathcal{B}(\pmb{P},J,J_{z})\rangle=&\sum_{m_{L},S_{z};c_{i}}\langle{L,m_{L}};{S,S_{z}}\vert{J,J_{z}}\rangle
{\int}d\pmb{p}_{1}d\pmb{p}_{2}d\pmb{p}_{3}\\
&\delta^{3}(\pmb{P}-\pmb{p}_{1}-\pmb{p}_{2}-\pmb{p}_{3})\Psi_{c;L,M_{L}}(\pmb{p}_1,\pmb{p}_2,\pmb{p}_3)\\
&\chi_{c;S,S_{z}}(s_1,s_2,s_3)\frac{\epsilon_{c_1c_2c_3}}{\sqrt{6}}\phi_{i_1,i_2,i_3}\\
&b^{\dagger}_{s_1,i_1,c_1}(\pmb{p}_1)b^{\dagger}_{s_2,i_2,c_2}(\pmb{p}_2)b^{\dagger}_{s_3,i_3,c_3}(\pmb{p}_3)\vert{0}\rangle
\end{split}
\end{equation}
with the normalization
\begin{equation}
\langle\mathcal{B}(\mathbf{P}^{\prime},J,J_{z})\vert\mathcal{B}(\mathbf{P},J,J_{z})\rangle=\delta^{3}(\mathbf{P}-\mathbf{P}^{\prime}).
\end{equation}
Here, $\pmb{P}$ represents the three-momentum of the baryon, and $(\pmb{p}_{1}, \pmb{p}_{2}, \pmb{p}_{3})$ are the three-momenta of the constituent quarks. Additionally, $\chi_{S,S_{z}}(s_1,s_2,s_3)$ is the spin wave function, and $\phi_{i_1,i_2,i_3}$ is the flavor wave function. The spatial wave function, $\Psi_{L,M_{L}}(\pmb{p}_1,\pmb{p}_2,\pmb{p}_3)$, is also defined and can be written as
\begin{equation}
\begin{split}
\Psi_{L,M_{L}}(\pmb{p}_1,\pmb{p}_2,\pmb{p}_3)=&\sum_{m_{\rho},m_{\lambda}}\langle{l_{\rho},m_{\rho};l_{\lambda},m_{\lambda}}\vert{L,M_{L}}\rangle\\
&\times\psi_{l_{\rho},m_{\rho}}(\pmb{\rho})\psi_{l_{\lambda},m_{\lambda}}(\pmb{\lambda}).
\label{eq:wavefunction}
\end{split}
\end{equation}

Following the approach in Ref.~\cite{Zeng:2022egh}, within the framework of NRQM, the nonperturbative parameters $f_{1}^{\mathcal{B}^{\prime}\mathcal{B}}$ and $g_{1}^{\mathcal{B}^{\prime}\mathcal{B}}$ can be expressed as momentum integrals of the baryon wave functions, given by
\begin{equation}
\begin{split}
f_{1}^{\mathcal{B}^{\prime}\mathcal{B}}=&(-1)\times\int{d\mathbf{p}_{1}d\mathbf{p}_{2}d\mathbf{p}_{3}d\mathbf{p}_{4}d\mathbf{p}_{5}d\mathbf{p}_{6}}{d\mathbf{p}_{i}d\mathbf{p}_{j}}\\
&\times\delta^{3}(\mathbf{p}_{1}+\mathbf{p}_{2}+\mathbf{p}_{3}-\mathbf{P}_{i})\delta^{3}(\mathbf{p}_{4}+\mathbf{p}_{5}+\mathbf{p}_{6}-\mathbf{P}_{f})\\
&\times\Psi_{\mathcal{B}^{\prime}}^{*}(\mathbf{p}_{4},\mathbf{p}_{5},\mathbf{p}_{6})\Psi_{\mathcal{B}}(\mathbf{p}_{1},\mathbf{p}_{2},\mathbf{p}_{3})\delta^{3}(\mathbf{p}_{i}-\mathbf{p}_{j})\\
&\times\langle\mathcal{B}^{\prime}{\!}\uparrow\vert{b_{q_{i}}^{\dagger}b_{q_{j}}}\vert\mathcal{B}{\!}\uparrow\rangle
\langle0\vert{b_{6}b_{5}b_{4}}{b_{i}^{\dagger}b_{j}}{b_{1}^{\dagger}b_{2}^{\dagger}b_{3}^{\dagger}}\vert0\rangle,
\label{eq:ffs_f1}
\end{split}
\end{equation}
and
\begin{equation}
\begin{split}
g_{1}^{\mathcal{B}^{\prime}\mathcal{B}}=&(-1)\times\int{d\mathbf{p}_{1}d\mathbf{p}_{2}d\mathbf{p}_{3}d\mathbf{p}_{4}d\mathbf{p}_{5}d\mathbf{p}_{6}}{d\mathbf{p}_{i}d\mathbf{p}_{j}}\\
&\times\delta^{3}(\mathbf{p}_{1}+\mathbf{p}_{2}+\mathbf{p}_{3}-\mathbf{P}_{i})\delta^{3}(\mathbf{p}_{4}+\mathbf{p}_{5}+\mathbf{p}_{6}-\mathbf{P}_{f})\\
&\times\Psi_{\mathcal{B}^{\prime}}^{*}(\mathbf{p}_{4},\mathbf{p}_{5},\mathbf{p}_{6})\Psi_{\mathcal{B}}(\mathbf{p}_{1},\mathbf{p}_{2},\mathbf{p}_{3})\delta^{3}(\mathbf{p}_{i}-\mathbf{p}_{j})\\
&\times\langle\mathcal{B}^{\prime}{\!}\uparrow\vert{b_{q_{i}}^{\dagger}b_{q_{j}}\sigma_{z}}\vert\mathcal{B}{\!}\uparrow\rangle
\langle0\vert{b_{6}b_{5}b_{4}}{b_{i}^{\dagger}b_{j}}{b_{1}^{\dagger}b_{2}^{\dagger}b_{3}^{\dagger}}\vert0\rangle,
\label{eq:ffs_g1}
\end{split}
\end{equation}
respectively. Similarly, the axial-vector form factor $g_{\mathcal{B}^{\prime}\mathcal{B}}^{A}$ can also be expressed as momentum integrals of the baryon wave functions, given by
\begin{equation}
\begin{split}
g_{\mathcal{B}^{\prime}\mathcal{B}}^{A}=&(-1)\times\int{d\mathbf{p}_{1}d\mathbf{p}_{2}d\mathbf{p}_{3}d\mathbf{p}_{4}d\mathbf{p}_{5}d\mathbf{p}_{6}}{d\mathbf{p}_{i}d\mathbf{p}_{j}}\\
&\times\delta^{3}(\mathbf{p}_{1}+\mathbf{p}_{2}+\mathbf{p}_{3}-\mathbf{P}_{i})\delta^{3}(\mathbf{p}_{4}+\mathbf{p}_{5}+\mathbf{p}_{6}-\mathbf{P}_{f})\\
&\times\Psi_{\mathcal{B}^{\prime}}^{*}(\mathbf{p}_{4},\mathbf{p}_{5},\mathbf{p}_{6})\Psi_{\mathcal{B}}(\mathbf{p}_{1},\mathbf{p}_{2},\mathbf{p}_{3})\delta^{3}(\mathbf{p}_{i}-\mathbf{p}_{j})\\
&\times\langle\mathcal{B}^{\prime}{\!}\uparrow\vert{b_{q_{i}}^{\dagger}b_{q_{j}}\sigma_{z}}\vert\mathcal{B}{\!}\uparrow\rangle
\langle0\vert{b_{6}b_{5}b_{4}}{b_{i}^{\dagger}b_{j}}{b_{1}^{\dagger}b_{2}^{\dagger}b_{3}^{\dagger}}\vert0\rangle.
\label{eq:g}
\end{split}
\end{equation}

In addition, the baryonic matrix elements $a_{\mathcal{B}^{\prime}\mathcal{B}}$ and $\tilde{a}_{\mathcal{B}^{\prime}\mathcal{B}}$ can be evaluated by
\begin{equation}
\begin{split}
a_{\mathcal{B}^{\prime}\mathcal{B}}=&\frac{G_{F}}{2\sqrt{2}}c_{-}V_{us}V_{ud}^{*}\langle\mathcal{B}^{\prime}\vert{O_{-}^{\text{PC}}}\vert\mathcal{B}\rangle,\\
\tilde{a}_{\mathcal{B}^{\prime}\mathcal{B}}=&\frac{G_{F}}{2\sqrt{2}}c_{-}V_{cs}V_{cd}^{*}\langle\mathcal{B}^{\prime}\vert{\tilde{O}_{-}^{\text{PC}}}\vert\mathcal{B}\rangle,
\end{split}
\end{equation}
respectively, with the operators $O_{-}=O_{1}-O_{2}$ and $\tilde{O}_{-}=\tilde{O}_{1}-\tilde{O}_{2}$, the Wilson coefficient $c_{-}=c_{1}-c_{2}$, and the integration being as
\begin{widetext}
\begin{equation}
\begin{split}
\langle\mathcal{B}^{\prime}\vert(\bar{q}_{i}q_{j})(\bar{q}_{k}q_{l})\vert\mathcal{B}\rangle=&\frac{1}{(2\pi)^{3}}\int{d\mathbf{p}_{1}d\mathbf{p}_{2}d\mathbf{p}_{3}d\mathbf{p}_{4}d\mathbf{p}_{5}d\mathbf{p}_{6}}{d\mathbf{p}_{i}d\mathbf{p}_{j}d\mathbf{p}_{k}d\mathbf{p}_{l}}
\Psi_{\mathcal{B}^{\prime}}^{*}(\mathbf{p}_{4},\mathbf{p}_{5},\mathbf{p}_{6})\Psi_{\mathcal{B}}(\mathbf{p}_{1},\mathbf{p}_{2},\mathbf{p}_{3})\\
&\times\delta^{3}(\mathbf{p}_{1}+\mathbf{p}_{2}+\mathbf{p}_{3}-\mathbf{P}_{i})\delta^{3}(\mathbf{p}_{4}+\mathbf{p}_{5}+\mathbf{p}_{6}-\mathbf{P}_{f})\delta^{3}(\mathbf{p}_{i}+\mathbf{p}_{k}-\mathbf{p}_{j}-\mathbf{p}_{l})\\
&\times6\langle\mathcal{B}^{\prime}{\!}\uparrow\vert(b^{\dagger}_{q_{i}}b_{q_{j}})_{1}(b^{\dagger}_{q_{k}}b_{q_{l}})_{2}(1-\pmb{\sigma}_{1}\cdot\pmb{\sigma}_{2})\vert\mathcal{B}{\!}\uparrow\rangle
\langle0\vert{b_{6}b_{5}b_{4}}{b_{i}^{\dagger}b_{j}}{b_{k}^{\dagger}b_{l}}{b_{1}^{\dagger}b_{2}^{\dagger}b_{3}^{\dagger}}\vert0\rangle.
\label{eq:a}
\end{split}
\end{equation}
\end{widetext}
It should be noted that the subscripts $1$ and $2$ indicate that the quark operators act only on the first and second quarks, respectively. To proceed with the calculation, we can expand $\pmb{\sigma}_{1}\cdot\pmb{\sigma}_{2}$ as
\begin{equation}
\pmb{\sigma}_{1}\cdot\pmb{\sigma}_{2}=\frac{1}{2}(\sigma_{1+}\sigma_{2-}+\sigma_{1-}\sigma_{2+})+\sigma_{1z}\sigma_{2z},
\end{equation}
where $\sigma_{\pm}=\sigma_{x}\pm i\sigma_{y}$.

\section{The nonrelativistic potential and baryon wave functions}
\label{sec3}

In this section, we employ the nonrelativistic quark model Hamiltonian to describe a baryon state, as outlined in Refs.~\cite{Copley:1979wj,Pervin:2007wa,Roberts:2007ni,Yoshida:2015tia}:
\begin{equation}
\mathcal{H}=\sum_{i=1,2,3}\Big{(}m_{i}+\frac{p_{i}^{2}}{2m_{i}}\Big{)}+\sum_{i<j}V_{ij},
\end{equation}
where $m_{i}$ and $p_{i}$ denote the mass and momentum of the $i$-th quark, respectively. The nonrelativistic potential is given by: $V_{ij} = V_{ij}^{\text{con}} + V_{ij}^{\text{short}}$ with the linear potential
\begin{equation}
V_{ij}^{\text{con}}=\frac{br_{ij}}{2}+\text{const.},
\end{equation}
and the short-range potential:
\begin{equation}
\begin{split}
V_{ij}^{\text{short}}=&-\frac{2}{3}\frac{\alpha^{\text{Coul}}}{r_{ij}}
+\frac{16\pi\alpha^{ss}}{9m_{i}m_{j}}\mathbf{S}_{i}\cdot\mathbf{S}_{j}\frac{\Lambda^{2}}{4\pi r_{ij}}\text{exp}(-\Lambda r_{ij})\\
&+\frac{\alpha^{\text{so}}\big{[}1{\!}-{\!}\text{exp}(-\Lambda r_{ij})\big{]}^{2}}{3r_{ij}^{3}}
\Bigg{[}\mathbf{L}_{ij}\cdot(\mathbf{S}_{i}+\mathbf{S}_{j})\\
&\times\Bigg{(}\frac{1}{m_{i}^{2}}+\frac{1}{m_{j}^{2}}+\frac{4}{m_{i}m_{j}}\Bigg{)}+\mathbf{L}_{ij}\cdot(\mathbf{S}_{i}-\mathbf{S}_{j})\Bigg{(}\frac{1}{m_{i}^{2}}-\frac{1}{m_{j}^{2}}\Bigg{)}\Bigg{]}\\
&+\frac{2\alpha^{\text{ten}}\big{[}1{\!}-{\!}\text{exp}(-\Lambda r_{ij})\big{]}^{2}}{3m_{i}m_{j}r_{ij}^{3}}
\Bigg{[}\frac{3(\mathbf{s}_{i}\cdot\mathbf{r}_{ij})(\mathbf{s}_{j}\cdot\mathbf{r}_{ij})}{r_{ij}^{2}}-\mathbf{S}_{i}\cdot\mathbf{S}_{j}\Bigg{]},
\end{split}
\end{equation}
where  $\alpha^{\text{Coul}}=K(m_{i}+m_{j})/(m_{i}m_{j})$, $\pmb{S}_{i}$ is the spin operator of the $i$-th quark, and $\pmb{L}_{ij}$ is the orbital operator. The parameters of the nonrelativistic potential are summarized in Table~\ref{tab:potential parameters}.

\begin{table}[htbp]
\centering
\caption{The parameters of the nonrelativistic potential~\cite{Yoshida:2015tia}. Additionally, the quark masses are taken as $m_{u,d}=300~\text{MeV}$, $m_{s}=510~\text{MeV}$, and $m_{c}=1750~\text{MeV}$.}
\label{tab:potential parameters}
\renewcommand\arraystretch{1.15}
\begin{tabular*}{83mm}{c@{\extracolsep{\fill}}ccc}
\toprule[1pt]
\toprule[0.5pt]
Parameters              &Values    &Parameters  &Values\\
\midrule[0.5pt]
$b~(\text{GeV}^{2})$     &$0.165$   &$\alpha^{\text{ss}}$  &$1.2$\\
$\text{const.}~(\text{GeV})$  &$-1.139$  &$\alpha^{\text{so}}=\alpha^{\text{ten}}$  &$0.077$\\
$K~(\text{MeV})$         &$90$      &$\Lambda~(\text{fm}^{-1})$  &$3.5$\\
\bottomrule[0.5pt]
\bottomrule[1pt]
\end{tabular*}
\end{table}

In the study of the baryon spectrum, the masses and wave functions can be obtained by solving the Schr\"{o}dinger equation
\begin{equation}
\mathcal{H}\vert\Psi_{J,M_{J}}\rangle=E\vert\Psi_{J,M_{J}}\rangle
\label{eq:Schrodinger}
\end{equation}
using the Rayleigh-Ritz variational principle. The baryon wave function $\Psi_{J,M_{J}}$ is expressed as a combination of color, spin, spatial, and flavor terms:
\begin{equation}
\begin{split}
\Psi_{J,M_{J}}=&\sum_{\alpha}C^{(\alpha)}\Psi_{J,M_{J}}^{(\alpha)},\\
\Psi_{J,M_{J}}^{(\alpha)}=&\chi^{\text{color}}\Big{\{}\chi_{S,M_{S}}^{\text{spin}}\psi_{L,M_{L}}^{\text{spatial}}\Big{\}}_{J,M_{J}}\psi^{\text{flavor}},
\end{split}
\end{equation}
where $C^{(\alpha)}$ is the coefficient, and $\alpha$ represents all possible quantum numbers. The spatial wave function $\psi_{L,M_{L}}^{\text{spatial}}$ consists of both $\rho$-mode and $\lambda$-mode excitations:
\begin{equation}
\psi_{L,M_{L}}^{\text{spatial}}(\pmb{\rho},\pmb{\lambda})=\sum_{m_{\rho},m_{\lambda}}\langle{l_{\rho},m_{\rho};l_{\lambda},m_{\lambda}}\vert{L,M_{L}}\rangle\psi_{l_{\rho},m_{\rho}}(\pmb{\rho}_{c})\psi_{l_{\lambda},m_{\lambda}}(\pmb{\lambda}_{c}),
\label{eq:wavefunction2}
\end{equation}
where the momentums of the $\rho$-mode and $\lambda$-mode are defined as
\begin{equation}
\begin{split}
\pmb{\rho}=&\frac{m_{1}\pmb{p}_{2}-m_{2}\pmb{p}_{1}}{m_{1}+m_{2}},\\
\pmb{\lambda}=&\frac{(m_{1}+m_{2})\pmb{p}_{3}-m_{3}(\pmb{p}_{1}+\pmb{p}_{2})}{m_{1}+m_{2}+m_{3}},
\end{split}
\end{equation}
respectively. This formulation assumes that the single heavy baryon can be treated as a bound state of a light quark cluster and a heavy quark, while the double heavy baryon is considered a bound state of a heavy quark cluster and a light quark.

With the nonrelativistic potential and baryon wave function prepared, we solve the Schr\"{o}dinger equation using the Gaussian expansion method~\cite{Hiyama:2003cu,Hiyama:2018ivm}. The infinitesimally-shift Gaussian basis~\cite{Hiyama:2003cu,Hiyama:2018ivm}
is given by 
\begin{equation}
\begin{split}
\phi_{nlm}^{G}(\mathbf{r})=&\phi^{G}_{nl}(r)~Y_{lm}(\pmb{\hat{r}})\\
=&\sqrt{\frac{2^{l+2}(2\nu_{n})^{l+3/2}}{\sqrt{\pi}(2l+1)!!}}\lim_{\varepsilon\rightarrow0}
\frac{1}{(\nu_{n}\varepsilon)^l}\sum_{k=1}^{k_{\text{max}}}C_{lm,k}e^{-\nu_{n}\big{(}\mathbf{r}-\varepsilon\vec{D}_{lm,k}\big{)}^2},
\label{eq:Gaussianbasis}
\end{split}
\end{equation}
which is used to expand the spatial wave functions $\psi_{l_{\rho},m_{\rho}}$ and $\psi_{l_{\lambda},m_{\lambda}}$. The Gaussian size parameter $\nu_{n}$ is chosen to follow a geometric progression as~\cite{Li:2021qod,Luo:2022cun}
\begin{equation}
\nu_{n}=1/r^2_{n},~~~r_{n}=r_{min}~a^{n-1},~~~a=\Big{(}r_{max}/r_{min}\Big{)}^{\frac{1}{n_{max}-1}}.
\end{equation}
Specifically, we set $r_{\rho_{min}}=0.2~\text{fm}$ and $r_{\rho_{max}}=2.0~\text{fm}$, and $n_{\rho_{max}}=6$, with the same Gaussian size parameters applied to the $\lambda$-mode: $r_{\lambda_{min}}=0.2~\text{fm}$, $r_{\lambda_{max}}=2.0~\text{fm}$ and $n_{\lambda_{max}}=6$. The Gaussian basis in momentum space is obtained by replacing $r\to p$ and $\nu_{n}\to1/(4\nu_{n})$.

Finally, we can obtain the baryon spatial wave functions and the masses
by solving the Schr\"{o}dinger equation:
\begin{equation}
\Big{(}T^{\alpha^{\prime},\alpha}+V^{\alpha^{\prime},\alpha}\Big{)}C^{(\alpha)}=EN^{\alpha^{\prime},\alpha}C^{(\alpha)},
\end{equation}
where the matrix elements are given by
\begin{equation}
\begin{split}
T^{\alpha^{\prime},\alpha}=&\Big{\langle}\Psi_{\mathbf{J},\mathbf{M_J}}^{(\alpha^{\prime})}\big{\vert}
\Bigg{[}\sum_{i=1,2,3}\Big{(}m_{i}+\frac{p_{i}^{2}}{2m_{i}}\Big{)}\Bigg{]}
\big{\vert}\Psi_{\mathbf{J},\mathbf{M_J}}^{(\alpha)}\Big{\rangle},\\
V^{\alpha^{\prime},\alpha}=&\Big{\langle}\Psi_{\mathbf{J},\mathbf{M_J}}^{(\alpha^{\prime})}\big{\vert}
\sum_{i<j}V_{ij}
\big{\vert}\Psi_{\mathbf{J},\mathbf{M_J}}^{(\alpha)}\Big{\rangle},\\
N^{\alpha^{\prime},\alpha}=&\Big{\langle}\Psi_{\mathbf{J},\mathbf{M_J}}^{(\alpha^{\prime})}
\big{\vert}\Psi_{\mathbf{J},\mathbf{M_J}}^{(\alpha)}\Big{\rangle}.
\label{eq:matrixelement}
\end{split}
\end{equation}
In Table~\ref{tab:wavefunctions}, we present the calculated masses of the single charmed baryons and the corresponding coefficients of the Gaussian bases. Comparing with the experimental measurements \cite{ParticleDataGroup:2024cfk}, our results are in good agreement.

\begin{table*}[htbp]\centering
\caption{Comparison of theoretical calculations and experimental values for the masses of single charmed baryons. The coefficients of the Gaussian bases are presented in the fourth column, with the pairs $(n_{\rho},n_{\lambda})$  arranged in the sequence $\{(1,1),(1,2),\cdots,(1,n_{\lambda_{max}}),(2,1),(2,2), \cdots,(2,n_{\lambda_{max}}),$ $\cdots,(n_{\rho_{max}},1), (n_{\rho_{max}},2),\cdots,(n_{\rho_{max}},n_{\lambda_{max}})\}$.}
\label{tab:wavefunctions}
\renewcommand\arraystretch{1.05}
\begin{tabular*}{172mm}{c@{\extracolsep{\fill}}ccc}
\toprule[1pt]
\toprule[0.5pt]
Charmed  &Theoretical values  &Experimantal values &\multirow{2}*{Eigenvector coefficients $C^{(\alpha)}$}\\
baryons  &$\text{(GeV)}$  &$\text{(MeV)}$~\cite{ParticleDataGroup:2024cfk}  &\\
\midrule[0.5pt]
\multirow{5}*{\shortstack{$\Lambda_c^{+}$}}  &\multirow{5}*{$2.286$} &\multirow{5}*{$2286.46\pm0.14$}
&$\big{\{}-0.0000, -0.0020, -0.0094, -0.0235, 0.0006, -0.0002, -0.0073,$\\
&&&$0.0130, -0.0213, -0.0227, -0.0002, 0.0000, 0.0041, -0.0398,$\\
&&&$-0.0100, -0.0913, 0.0052, -0.0017, -0.0010, 0.0229, -0.2326,$\\
&&&$-0.3155, 0.0106, -0.0028, -0.0043, 0.0159, -0.0376, -0.4100,$\\
&&&$0.0089, -0.0036, 0.0012, -0.0042, 0.0089, 0.0333, -0.0049, 0.0012\big{\}}$\\
\specialrule{0em}{2pt}{2pt}
\multirow{5}*{\shortstack{$\Sigma_{c}^{++}/\Sigma_{c}^{+}/\Sigma_{c}^{0}$}} &\multirow{5}*{$2.467$} &\multirow{5}*{\makecell[c]{$2453.97\pm0.14$\\$2452.65^{+0.22}_{-0.16}$\\$2453.75\pm0.14$}}
&$\big{\{}0.0016, -0.0041, 0.0056, 0.0018, 0.0002, -0.0000, -0.0114,$\\
&&&$0.0270, -0.0303, 0.0011, -0.0028, 0.0008, 0.0069, -0.0653,$\\
&&&$0.0895, 0.0033, 0.0067, -0.0021, 0.0028, 0.0188, -0.2988,$\\
&&&$-0.1571, 0.0007, 0.0004, -0.0092, 0.0352, -0.0780, -0.6505,$\\
&&&$0.0240, -0.0081, 0.0026, -0.0096, 0.0198, 0.0365, -0.0108, 0.0026\big{\}}$\\
\specialrule{0em}{2pt}{2pt}
\multirow{5}*{\shortstack{$\Xi_c^{+}/\Xi_{c}^{0}$}} &\multirow{5}*{$2.500$} &\multirow{5}*{\makecell[c]{$2467.71\pm0.23$\\$2470.44\pm0.28$}}
&$\big{\{}-0.0005, -0.0006, -0.0104, -0.0137, 0.0009, -0.0002, -0.0059,$\\
&&&$0.0089, -0.0168, -0.0243, 0.0028, -0.0007, 0.0024, -0.0353,$\\
&&&$-0.0271, -0.0476, 0.0035, -0.0008, -0.0030, 0.0266, -0.3020,$\\
&&&$-0.3543, 0.0370, -0.0083, -0.0042, 0.0200, -0.0665, -0.3414,$\\
&&&$0.0273, -0.0064, 0.0008, -0.0036, 0.0100, 0.0351, -0.0046, 0.0011\big{\}}$\\
\specialrule{0em}{2pt}{2pt}
\multirow{5}*{\shortstack{$\Xi_c^{\prime+}/\Xi_{c}^{\prime0}$}} &\multirow{5}*{$2.603$} &\multirow{5}*{\makecell[c]{$2578.2\pm0.5$\\$2578.7\pm0.5$}}
&$\big{\{}0.0006, -0.0015, 0.0021, 0.0038, -0.0009, 0.0002, -0.0094,$\\
&&&$0.0181, -0.0181, -0.0128, 0.0017, -0.0004, 0.0043, -0.0575,$\\
&&&$0.0569, 0.0325, -0.0037, 0.0008, -0.0003, 0.0193, -0.3614,$\\
&&&$-0.2730, 0.0295, -0.0064, -0.0072, 0.0347, -0.1110, -0.4803,$\\
&&&$0.0435, -0.0101, 0.0014, -0.0063, 0.0171, 0.0439, -0.0072, 0.0017\big{\}}$\\
\specialrule{0em}{2pt}{2pt}
\multirow{5}*{\shortstack{$\Omega_{c}^{0}$}}  &\multirow{5}*{$2.737$} &\multirow{5}*{$2695.2\pm1.7$}
&$\big{\{}0.0018, -0.0043, 0.0035, 0.0025, -0.0004, 0.0001, -0.0157,$\\
&&&$0.0323, -0.0283, -0.0087, -0.0007, 0.0003, 0.0103, -0.0905,$\\
&&&$0.0648, 0.01720, 0.0027, -0.0010, -0.0020, 0.0448, -0.4914,$\\
&&&$-0.3327, 0.0391, -0.0079, -0.0086, 0.0324, -0.0679, -0.3229,$\\
&&&$0.0451, -0.0103, 0.0020, -0.0069, 0.0107, 0.0409, -0.0074, 0.0018\big{\}}$\\
\toprule[0.5pt]
\toprule[1pt]
\end{tabular*}
\end{table*}

\section{Numerical results}
\label{sec4}

Using the obtained baryon spatial wave functions as inputs, we can compute the numerical results of nonperturbative parameters using Eqs.~(\ref{eq:ffs_f1}-\ref{eq:g}) and Eq.~\eqref{eq:a}. The coefficients $\langle\mathcal{B}^{\prime}{\!}\uparrow\vert\cdots\vert\mathcal{B}{\!}\uparrow\rangle$, which depend on the spin-flavor wave functions of the baryons, along with the corresponding numerical results for the nonperturbative parameters, are presented in Tables~\ref{tab:nonperturbativeparameters1}, \ref{tab:nonperturbativeparameters2}, and \ref{tab:nonperturbativeparameters3}.

\begin{table}[htbp]
\centering
\caption{Coefficients $\langle\mathcal{B}^{\prime}\uparrow\vert\cdots\vert\mathcal{B}\uparrow\rangle$ and calculated values of nonperturbative parameters $f_{1}^{\mathcal{B}^{\prime}\mathcal{B}}$ and $g_{1}^{\mathcal{B}^{\prime}\mathcal{B}}$.}
\label{tab:nonperturbativeparameters1}
\renewcommand\arraystretch{1.05}
\begin{tabular*}{86mm}{c@{\extracolsep{\fill}}ccccc}
\toprule[1pt]
\toprule[0.5pt]
Parameters  &Coefficients  &Values  &Parameters  &Coefficients  &Values\\
\midrule[0.5pt]
$f_{1}^{\Lambda_{c}^{+}\Xi_{c}^{\bar{3},0}}$   &$1$    &$0.988$  &$g_{1}^{\Lambda_{c}^{+}\Xi_{c}^{\bar{3},0}}$   &$0$  &$0$\\
$f_{1}^{\Lambda_{c}^{+}\Xi_{c}^{6,0}}$         &$0$    &$0$   &$g_{1}^{\Lambda_{c}^{+}\Xi_{c}^{6,0}}$        &$\sqrt{\frac{1}{3}}$  &$0.562$\\
$f_{1}^{\Lambda_{c}^{+}\Xi_{c}^{\bar{3},+}}$   &$-1$   &$-0.988$    &$g_{1}^{\Lambda_{c}^{+}\Xi_{c}^{\bar{3},+}}$   &$0$  &$0$\\
$f_{1}^{\Lambda_{c}^{+}\Xi_{c}^{6,+}}$         &$0$    &$0$  &$g_{1}^{\Lambda_{c}^{+}\Xi_{c}^{6,+}}$        &$-\sqrt{\frac{1}{3}}$   &$-0.562$\\
$f_{1}^{\Xi_{c}^{\bar{3},+}\Omega_{c}^{0}}$    &$0$       &$0$  &$g_{1}^{\Xi_{c}^{\bar{3},+}\Omega_{c}^{0}}$     &$-\sqrt{\frac{2}{3}}$  &$-0.803$\\
$f_{1}^{\Xi_{c}^{6,+}\Omega_{c}^{0}}$          &$\sqrt{2}$   &$1.387$      &$g_{1}^{\Xi_{c}^{6,+}\Omega_{c}^{0}}$    &$\frac{2\sqrt{2}}{3}$  &$0.925$\\
$f_{1}^{\Xi_{c}^{\bar{3},0}\Omega_{c}^{0}}$    &$0$       &$0$  &$g_{1}^{\Xi_{c}^{\bar{3},0}\Omega_{c}^{0}}$     &$-\sqrt{\frac{2}{3}}$  &$-0.803$\\
$f_{1}^{\Xi_{c}^{6,0}\Omega_{c}^{0}}$          &$\sqrt{2}$   &$1.387$    &$g_{1}^{\Xi_{c}^{6,0}\Omega_{c}^{0}}$      &$\frac{2\sqrt{2}}{3}$  &$0.925$\\
\bottomrule[0.5pt]
\bottomrule[1pt]
\end{tabular*}
\end{table}

\begin{table}[htbp]
\centering
\caption{The coefficients $\langle\mathcal{B}^{\prime}\uparrow\vert\cdots\vert\mathcal{B}\uparrow\rangle$ and calculated values of nonperturbative parameters $g_{\mathcal{B}^{\prime}\mathcal{B}}^{A}$.}
\label{tab:nonperturbativeparameters2}
\renewcommand\arraystretch{1.15}
\begin{tabular*}{86mm}{c@{\extracolsep{\fill}}ccccc}
\toprule[1pt]
\toprule[0.5pt]
Parameters  &Coefficients  &Values  &Parameters  &Coefficients  &Values  \\
\midrule[0.5pt]
$g_{\Lambda_{c}^{+}\Sigma_{c}^{0}}^{A(\pi^{-})}$     &$\sqrt{\frac{2}{3}}$   &$0.795$  &$g_{\Xi_{c}^{\bar{3},+}\Xi_{c}^{6,0}}^{A(\pi^{-})}$   &$-\sqrt{\frac{1}{3}}$  &$-0.570$\\
$g_{\Xi_{c}^{6,+}\Xi_{c}^{\bar{3},0}}^{A(\pi^{-})}$  &$-\sqrt{\frac{1}{3}}$  &$-0.570$  &$g_{\Xi_{c}^{6,+}\Xi_{c}^{6,0}}^{A(\pi^{-})}$         &$\frac{2}{3}$  &$0.667$\\
$g_{\Lambda_{c}^{+}\Sigma_{c}^{+}}^{A(\pi^{0})}$     &$\sqrt{\frac{1}{3}}$   &$0.562$  &$g_{\Xi_{c}^{\bar{3},+}\Xi_{c}^{6,+}}^{A(\pi^{0})}$   &$-\frac{1}{2\sqrt{3}}$  &$-0.285$\\
$g_{\Xi_{c}^{6,+}\Xi_{c}^{\bar{3},+}}^{A(\pi^{0})}$  &$-\frac{1}{2\sqrt{3}}$  &$-0.285$  &$g_{\Xi_{c}^{6,+}\Xi_{c}^{6,+}}^{A(\pi^{0})}$         &$\frac{1}{3}$  &$0.333$\\
$g_{\Xi_{c}^{\bar{3},0}\Xi_{c}^{6,0}}^{A(\pi^{0})}$  &$\frac{1}{2\sqrt{3}}$  &$0.285$  &$g_{\Xi_{c}^{6,0}\Xi_{c}^{\bar{3},0}}^{A(\pi^{0})}$         &$\frac{1}{2\sqrt{3}}$  &$-0.285$\\
$g_{\Xi_{c}^{6,0}\Xi_{c}^{6,0}}^{A(\pi^{0})}$  &$-\frac{1}{3}$  &$-0.333$  &         &  &\\
\bottomrule[0.5pt]
\bottomrule[1pt]
\end{tabular*}
\end{table}

\begin{table*}[htbp]
\centering
\caption{The coefficients $\langle\mathcal{B}^{\prime}\uparrow\vert\cdots\vert\mathcal{B}\uparrow\rangle$ and calculated values of nonperturbative parameters $a_{\mathcal{B}^{\prime}\mathcal{B}}$ and $\tilde{a}_{\mathcal{B}^{\prime}\mathcal{B}}$.}
\label{tab:nonperturbativeparameters3}
\renewcommand\arraystretch{1.15}
\begin{tabular*}{110mm}{c@{\extracolsep{\fill}}ccccc}
\toprule[1pt]
\toprule[0.5pt]
Parameters  &Coefficients  &Values  &Parameters  &Coefficients  &Values\\
\midrule[0.5pt]
$a_{\Lambda_{c}^{+}\Xi_{c}^{\bar{3},+}}$  &$8$  &$0.0163G_{F}$  &$\tilde{a}_{\Lambda_{c}^{+}\Xi_{c}^{\bar{3},+}}$  &$2$    &$-0.00297G_{F}$\\
$\tilde{a}_{\Lambda_{c}^{+}\Xi_{c}^{6,+}}$        &$-2\sqrt{3}$  &$0.00497G_{F}$ 
&$\tilde{a}_{\Sigma_{c}^{0}\Xi_{c}^{\bar{3},0}}$   &$2\sqrt{6}$   &$-0.00661G_{F}$\\
$\tilde{a}_{\Sigma_{c}^{0}\Xi_{c}^{6,0}}$         &$-6\sqrt{2}$  &$0.0113G_{F}$
&$\tilde{a}_{\Sigma_{c}^{+}\Xi_{c}^{\bar{3},+}}$   &$2\sqrt{3}$   &$-0.00468G_{F}$\\
$\tilde{a}_{\Sigma_{c}^{+}\Xi_{c}^{6,+}}$         &$-6$  &$0.00801G_{F}$
&$\tilde{a}_{\Xi_{c}^{\bar{3},0}\Omega_{c}^{0}}$   &$-2\sqrt{6}$   &$0.00644G_{F}$\\
$\tilde{a}_{\Xi_{c}^{6,0}\Omega_{c}^{0}}$         &$-6\sqrt{2}$   &$0.0104G_{F}$ &&&\\
\bottomrule[0.5pt]
\bottomrule[1pt]
\end{tabular*}
\end{table*}

It should be mention that, the form factors $f_{1}$ and $g_{1}$ calculated by Eqs.~(\ref{eq:ffs_f1}-\ref{eq:ffs_g1}) are working at $(m_{i}-m_{f})^{2}$ point. The knowledge of the form factors being dependence on $q^{2}$, should be taken into account. However, considering $(m_{\Xi_{c}}-m_{\Lambda_{c}})^{2}\approx(m_{\Omega_{c}}-m_{\Xi_{c}})^{2}\approx m_{\pi}^{2}$, the calculated form factors can be used directly.

With the obtained nonperturbative parameters, the factorizable and nonfactorizable amplitudes for the relevant HFC processes can be calculated numerically. The decay width and the asymmetry parameter $\alpha$ are then determined by the following expressions:
\begin{equation}
\begin{split}
\Gamma=&\frac{\vert{p_{c}}\vert}{8\pi}\Bigg{[}\frac{(m_{i}+m_{f})^{2}-m_{P}^{2}}{m_{i}^{2}}\vert{A}\vert^{2}+\frac{(m_{i}-m_{f})^{2}-m_{P}^{2}}{m_{i}^{2}}\vert{B}\vert^{2}\Bigg{]},\\
\alpha=&\frac{2\kappa\text{Re}(A^{*}B)}{\vert{A}\vert^{2}+\kappa^{2}\vert{B}\vert^{2}},
\end{split}
\end{equation}
respectively, where $\kappa=\vert{p_{c}}\vert/(E_{f}+m_{f})$ is the kinematic factor, with $\vert{p_{c}}\vert$ being the momentum of the final baryon, and $E_{f}=\sqrt{p_{c}^{2}+m_{f}^{2}}$ being the corresponding energy. The lifetimes of charmed baryons are given by ~\cite{ParticleDataGroup:2024cfk}
\begin{equation}
\begin{split}
\tau_{\Xi_{c}^{0}}=&1.504\times10^{-13}~s,~\tau_{\Xi_{c}^{+}}=4.53\times10^{-13}~s,\\
\tau_{\Omega_{c}^{0}}=&2.73\times10^{-13}~s,
\label{eq:lifetimes}
\end{split}
\end{equation}
which is used to evaluate the corresponding branching fractions.

In our approach, the only free parameter is the mixing angle $\theta$. From mass relations, the mixing angle is approximately determined to be $\theta = \pm (24.7 \pm 0.9)^\circ$ without fixing the sign~\cite{Geng:2022yxb}. Additionally, the LHCb Collaboration has reported the measurement~\cite{LHCb:2020gge} 
\begin{equation*}
\mathcal{B}(\Xi_{c}^{0}\to\Lambda_{c}^{+}\pi^{-})=(0.55\pm0.02\pm0.18)\%,
\end{equation*}
and the Belle Collaboration reported~\cite{Belle:2022kqi}
\begin{equation*}
\mathcal{B}(\Xi_{c}^{0}\to\Lambda_{c}^{+}\pi^{-})=(0.54\pm0.05\pm0.05\pm0.12)\%.
\end{equation*}
The Particle Data Group fit the absolute branching fraction as $\mathcal{B}(\Xi_{c}^{0} \to \Lambda_{c}^{+} \pi^{-}) = (0.55 \pm 0.11)\%$~\cite{ParticleDataGroup:2024cfk}. This value serves as a constraint on the mixing angle $\theta$.

\begin{figure*}[htbp]\centering
  \begin{tabular}{cc}
  \includegraphics[width=60mm]{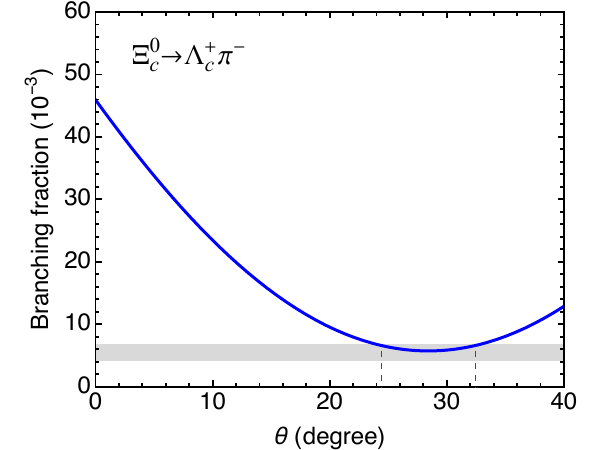}
  \includegraphics[width=60mm]{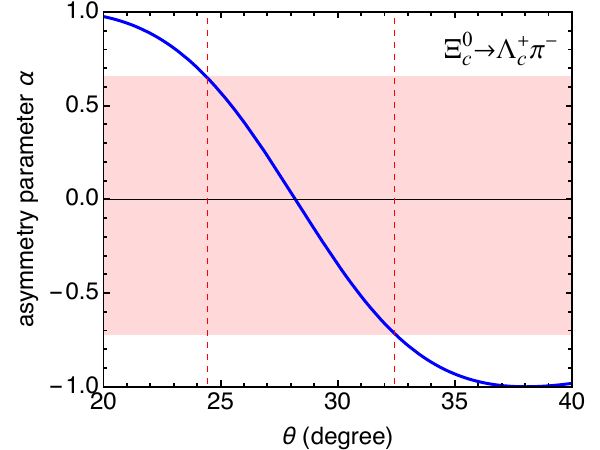}
  \end{tabular}
  \caption{The $\theta$ dependence of the branching fraction (top panel) and asymmetry parameter $\alpha$ (bottom panel) of $\Xi_{c}^{0}\to\Lambda_{c}^{+}\pi^{-}$ process.}
\label{fig:decay1}
\end{figure*}

In the left panel of Fig.~\ref{fig:decay1}, we show the dependence of the branching fraction $\mathcal{B}(\Xi_{c}^{0} \to \Lambda_{c}^{+} \pi^{-})$ on the mixing angle $\theta$. The blue curve represents our results, while the gray band indicates the experimental value. Clearly, the experimental value $\mathcal{B}(\Xi_{c}^{0} \to \Lambda_{c}^{+} \pi^{-}) = (0.55 \pm 0.11)\%$~\cite{ParticleDataGroup:2024cfk} is well reproduced for the mixing angle range $\theta \in (24.4^\circ, 32.4^\circ)$. Moreover, the minimal value of the branching fraction is found to be $\mathcal{B}(\Xi_{c}^{0} \to \Lambda_{c}^{+} \pi^{-}) = 5.73 \times 10^{-3}$ for $\theta = 28.4^\circ$, which closely approximates the central value of the experimental result. Additionally, we examine the asymmetry parameter $\alpha$ and show its dependence on $\theta$ in the right panel of Fig.~\ref{fig:decay1}. The light red region represents our predictions for $\theta \in (24.4^\circ, 32.4^\circ)$. It is evident that $\alpha$ is highly sensitive to the mixing angle, and thus, experimental measurements of this observable will provide a useful constraint on the mixing angle in future studies.

\begin{figure*}[htbp]\centering
  \begin{tabular}{lr}
  \includegraphics[width=60mm]{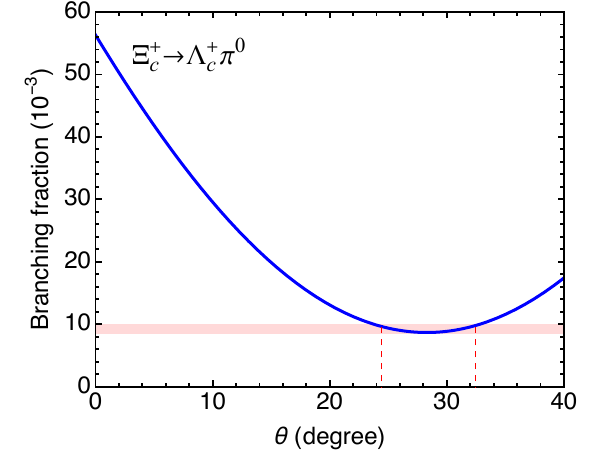}
  \includegraphics[width=60mm]{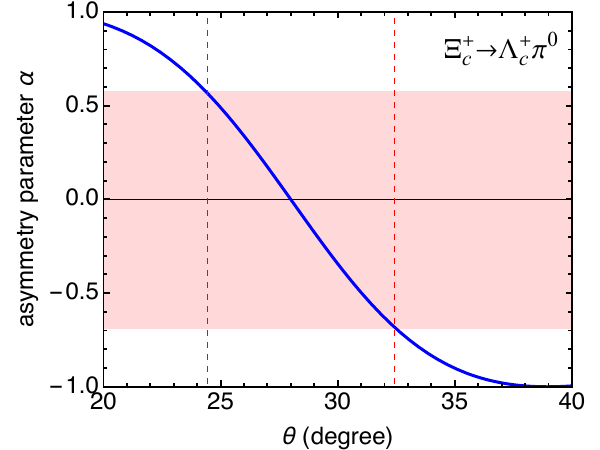}
  \end{tabular}
  \caption{The $\theta$ dependence of the branching fraction (left panel) and asymmetry parameter $\alpha$ (right panel) of $\Xi_{c}^{+}\to\Lambda_{c}^{+}\pi^{0}$ process.}
\label{fig:decay2}
\end{figure*}

\begin{figure*}[htbp]\centering
  \begin{tabular}{lr}
  \includegraphics[width=60mm]{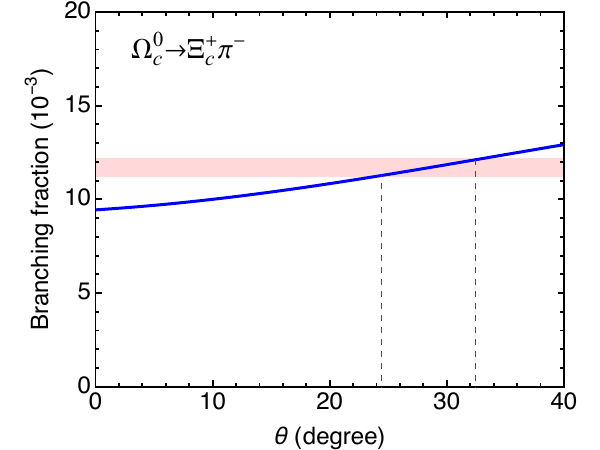}
  \includegraphics[width=60mm]{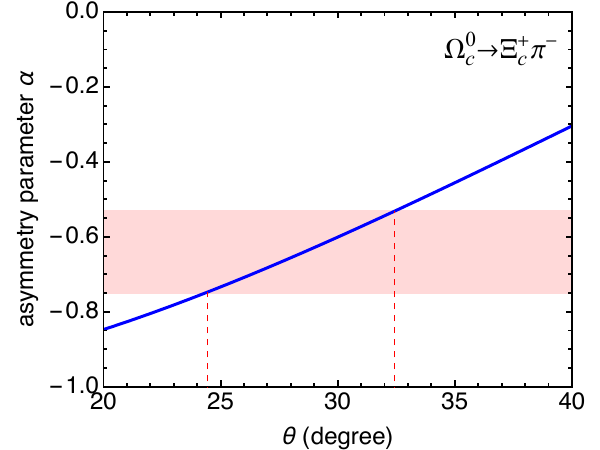}
  \end{tabular}
  \caption{The $\theta$ dependence of the branching fraction (left panel) and asymmetry parameter $\alpha$ (right panel) of $\Omega_{c}^{0}\to\Xi_{c}^{+}\pi^{-}$ process.}
\label{fig:decay3}
\end{figure*}

\begin{figure*}[htbp]\centering
  \begin{tabular}{lr}
  \includegraphics[width=60mm]{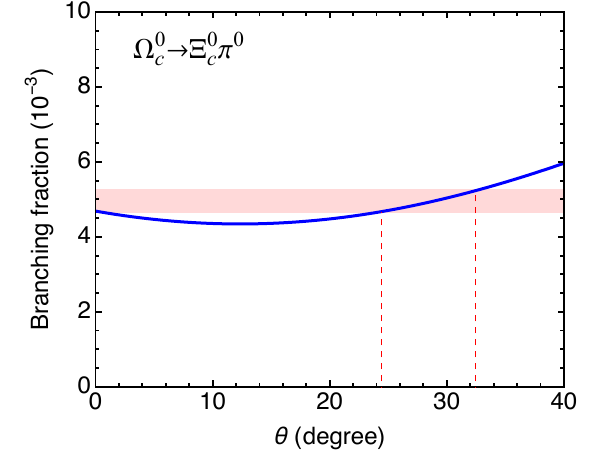}
  \includegraphics[width=60mm]{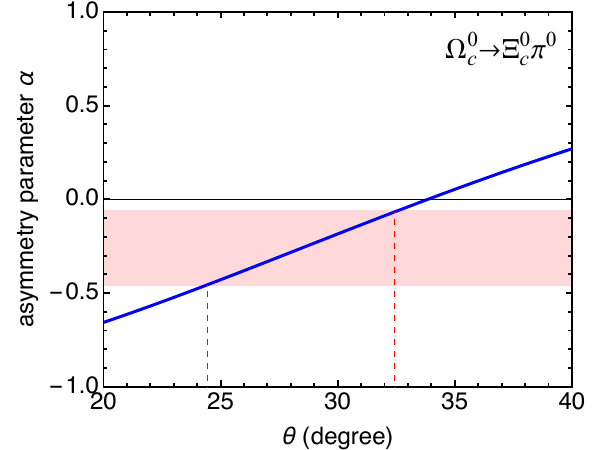}
  \end{tabular}
  \caption{The $\theta$ dependence of the branching fraction (left panel) and asymmetry parameter $\alpha$ (right panel) of $\Omega_{c}^{0}\to\Xi_{c}^{0}\pi^{0}$ process.}
\label{fig:decay4}
\end{figure*}

We also investigate the $\theta$ dependence of the branching fractions and asymmetry parameters of  the decays $\Xi_{c}^{+}\to\Lambda_{c}^{+}\pi^{0}$, $\Omega_{c}^{0}\to\Xi_{c}^{+}\pi^{-}$ and $\Omega_{c}^{0}\to\Xi_{c}^{0}\pi^{0}$,
presenting our results in Figs.~\ref{fig:decay2}, \ref{fig:decay3}, and \ref{fig:decay4}, respectively. For a mixing angle range $\theta \in (24.4^\circ, 32.4^\circ)$, we estimate the following branching fractions: 
\begin{equation*}
\begin{split}
\mathcal{B}(\Xi_{c}^{+}\to\Lambda_{c}^{+}\pi^{0})=&(8.69\sim9.79)\times10^{-3},\\
\mathcal{B}(\Omega_{c}^{0}\to\Xi_{c}^{+}\pi^{-})=&(11.3\sim12.1)\times10^{-3},\\
\mathcal{B}(\Omega_{c}^{0}\to\Xi_{c}^{0}\pi^{0})=&(4.67\sim5.23)\times10^{-3}.
\end{split}
\end{equation*}

Recently, the analyses by lattice QCD (LQCD) and QCDSR suggested a small mixing angle as $(1.2\pm0.1)\degree$~\cite{Liu:2023feb} and $(1.2\sim 2.8)\degree$~\cite{Sun:2023noo}, respectively. In our calculations, when the mixing angle is set to $\theta = 0\degree$, the branching fractions of the $\Xi_{c}^{0,+} \to \Lambda_{c}\pi^{-,0}$ processes reach up to the order of $10^{-2}$. This value is excessively large for Cabibbo-suppressed weak decays. By comparing our results with those in Refs.~\cite{Cheng:2022kea,Cheng:2022jbr}, we observe that the baryonic matrix elements obtained using the NRQM are approximately three times larger than those evaluated using the bag model. The large branching fractions calculated for the $\Xi_{c}^{0,+} \to \Lambda_{c}\pi^{-,0}$ processes at $\theta = 0\degree$ suggest that the baryonic matrix elements in this work may be overestimated, particularly if a small $\Xi_{c}-\Xi_{c}^{\prime}$ mixing angle is determined experimentally. To clarify this issue, further experimental determination of the mixing angle, as well as additional measurements of the physical observables in $\Xi_{c}^{0,+} \to \Lambda_{c}\pi^{-,0}$ decays, will be essential. Moreover, it is evident that our results for the branching fractions $\mathcal{B}(\Omega_{c}^{0}\to\Xi_{c}^{+,0}\pi^{-,0})$ exhibit significantly less sensitivity to the mixing angle $\theta$ compared to $\mathcal{B}(\Xi_{c}^{0,+}\to\Lambda_{c}\pi^{-,0})$. Therefore, future experimental measurements of the absolute branching fractions for $\Omega_{c}^{0}\to\Xi_{c}^{+,0}\pi^{-,0}$ will provide valuable insights for testing our predictions and determining whether the baryonic matrix elements are overestimated.

To further understand the relations between these decays, we employ flavor symmetry analysis, which is a powerful method for studying decay patterns~\cite{Groote:2021pxt}. Based on isospin SU(2) symmetry, the $\Delta I = 1/2$ rule suggests the following relations:
\begin{equation}
\begin{split}
\sqrt{2}\mathcal{M}(\Xi_{c}^{+}\to\Lambda_{c}^{+}\pi^{0})=&\mathcal{M}(\Xi_{c}^{0}\to\Lambda_{c}^{+}\pi^{-}),\\
\sqrt{2}\mathcal{M}(\Omega_{c}^{0}\to\Xi_{c}^{0}\pi^{0})=&\mathcal{M}(\Omega_{c}^{0}\to\Xi_{c}^{+}\pi^{-}).
\end{split}
\end{equation}
This leads to the following approximate branching fraction ratios
\begin{equation}
\frac{\mathcal{B}(\Xi_{c}^{0}\to\Lambda_{c}^{+}\pi^{-})}{\mathcal{B}(\Xi_{c}^{+}\to\Lambda_{c}^{+}\pi^{0})}\approx0.66,~~~
\frac{\mathcal{B}(\Omega_{c}^{0}\to\Xi_{c}^{+}\pi^{-})}{\mathcal{B}(\Omega_{c}^{0}\to\Xi_{c}^{0}\pi^{0})}=2,
\end{equation}
where the lifetimes of the charmed baryons are taken from Eq. \eqref{eq:lifetimes}. These results are in agreement with the $\Delta I = 1/2$ rule, and we expect that future measurements by the LHCb, Belle II, or BESIII experiments will provide a test of these predictions.

In Table~\ref{tab:amplitudes}, we present the numerical results for the amplitudes, branching fractions, and asymmetry parameters of the relevant HFC weak decays, with the mixing angle $\theta = 24.4^\circ$, $28.4^\circ$, and $32.4^\circ$. It is evident that the nonfactorizable PC amplitudes are highly sensitive to the mixing angle, which causes the asymmetry parameter $\alpha$ to exhibit erratic behavior. When referring back to Eqs.~(\ref{eq:Bnf1}-\ref{eq:Bnf4}), the high sensitivity of the calculated $B^{\text{nf}}$ to the mixing angle $\theta$ arises from the constructive or destructive interference of amplitudes originating from different intermediate charmed baryons. This sensitivity is further amplified by the factor $1/(m_{i,f}-m_{n})$ and the baryonic matrix element. The baryonic matrix element is crucial for reliable calculations, and more theoretical studies are needed to better understand its behavior. The amplitudes $B^{\text{nf}}$ for the $\Omega_{c}^{0}\to\Xi_{c}^{+,0}\pi^{-,0}$ processes exhibit less sensitivity to the mixing angle. Therefore, future experimental measurements of $\Omega_{c}^{0}\to\Xi_{c}^{+,0}\pi^{-,0}$ will be invaluable for verifying the suitability of the baryonic matrix element adopted in this work.

\begin{table*}[htbp]\centering
\caption{The factorizable and nonfactorizable amplitudes (in the unit of $10^{-2}G_{F}$), as well as the branching fractions and asymmetry parameter $\alpha$ of the concerned HFC processes, with the mixing angle $\theta=24.4\degree$, $28.4\degree$ and $32.4\degree$, respectively.}
\label{tab:amplitudes}
\renewcommand\arraystretch{1.15}
\begin{tabular*}{140mm}{c@{\extracolsep{\fill}}ccccccc}
\toprule[1pt]
\toprule[0.5pt]
Modes  &$\theta$  &$A^{\text{fac}}$  &$A^{\text{nf}}$  &$B^{\text{fac}}$  &$B^{\text{nf}}$  &$\mathcal{B}$  &$\alpha$\\
\midrule[0.5pt]
\multirow{3}*{\shortstack{$\Xi_{c}^{0}\to\Lambda_{c}^{+}\pi^{-}$}}
&$24.4\degree$  &$-0.416$  &$10.9$  &$2.78$  &$152$  &$6.60\times10^{-3}$  &$0.65$\\
&$28.4\degree$  &$-0.402$  &$10.8$  &$3.20$  &$-12.3$  &$5.74\times10^{-3}$  &$-0.04$\\
&$32.4\degree$  &$-0.386$  &$10.7$  &$3.60$  &$-177$  &$6.60\times10^{-3}$  &$-0.72$\\
\specialrule{0em}{2pt}{2pt}
\multirow{3}*{\shortstack{$\Xi_{c}^{+}\to\Lambda_{c}^{+}\pi^{0}$}}
&$24.4\degree$  &$-0.141$  &$7.72$  &$0.96$  &$96.8$  &$9.65\times10^{-3}$  &$0.57$\\
&$28.4\degree$  &$-0.137$  &$7.66$  &$1.10$  &$-12.7$  &$8.69\times10^{-3}$  &$-0.08$\\
&$32.4\degree$  &$-0.131$  &$7.57$  &$1.24$  &$-122$  &$9.79\times10^{-3}$  &$-0.68$\\
\specialrule{0em}{2pt}{2pt}
\multirow{3}*{\shortstack{$\Omega_{c}^{0}\to\Xi_{c}^{+}\pi^{-}$}}
&$24.4\degree$  &$-0.328$  &$-7.82$  &$-4.53$  &$110$  &$11.3\times10^{-3}$  &$-0.75$\\
&$28.4\degree$  &$-0.378$  &$-8.16$  &$-3.45$  &$93.2$  &$11.7\times10^{-3}$  &$-0.64$\\
&$32.4\degree$  &$-0.425$  &$-8.47$  &$-2.36$  &$76.1$  &$12.1\times10^{-3}$  &$-0.53$\\
\specialrule{0em}{2pt}{2pt}
\multirow{3}*{\shortstack{$\Omega_{c}^{0}\to\Xi_{c}^{0}\pi^{0}$}}
&$24.4\degree$  &$0.112$  &$5.53$  &$1.56$  &$-41.6$  &$4.67\times10^{-3}$  &$-0.46$\\
&$28.4\degree$  &$0.129$  &$5.77$  &$1.19$  &$-24.3$  &$4.92\times10^{-3}$  &$-0.26$\\
&$32.4\degree$  &$0.145$  &$5.99$  &$0.815$  &$-6.89$  &$5.23\times10^{-3}$  &$-0.07$\\
\bottomrule[0.5pt]
\bottomrule[1pt]
\end{tabular*}
\end{table*}

In Table~\ref{tab:Br}, we compare our numerical results for the branching fractions with experimental data and various theoretical predictions. Specifically, the authors of Ref.~\cite{Cheng:2015ckx} studied these decays within the frameworks of heavy-quark symmetry and chiral symmetry, calculating the relevant nonperturbative parameters using both the MIT bag model and the diquark model. Meanwhile, the authors of Refs.~\cite{Cheng:2022kea,Cheng:2022jbr,Niu:2021qcc,Ivanov:2023wir} employed the pole model, with parameters derived from NRQM~\cite{Niu:2021qcc}, the bag model and diquark model~\cite{Cheng:2022kea}, the bag model with the center-of-mass motion removed~\cite{Cheng:2022jbr}, and CCQM~\cite{Ivanov:2023wir}, respectively. Additionally, the study in Ref.~\cite{Gronau:2016xiq} concluded that the branching fractions are on the order of $10^{-3}$, assuming constructive interference between the amplitudes $A_{s\to u\bar{u}d}$ and $A_{cs\to cd}$, and on the order of $10^{-4}$ for destructive interference.

\begin{table*}[htbp]\centering
\caption{The comparison of different theoretical predictions and experimental data for the branching fractions of the relevant HFC weak decays.}
\label{tab:Br}
\renewcommand\arraystretch{1.25}
\begin{threeparttable}
\begin{tabular*}{160mm}{c@{\extracolsep{\fill}}ccccc}
\toprule[1pt]
\toprule[0.5pt]
Branching fractions  &$\Xi_{c}^{0}\to\Lambda_{c}^{+}\pi^{-}$  &$\Xi_{c}^{+}\to\Lambda_{c}^{+}\pi^{0}$ &$\Omega_{c}^{0}\to\Xi_{c}^{+}\pi^{-}$  &$\Omega_{c}^{0}\to\Xi_{c}^{0}\pi^{0}$\\
\midrule[0.5pt]
This work  &$(5.73\sim6.60)\times10^{-3}$  &$(8.69\sim9.79)\times10^{-3}$  &$(11.3\sim12.1)\times10^{-3}$  &$(4.67\sim5.23)\times10^{-3}$\\
Ref.~\cite{Cheng:2015ckx}  &$0.87\times10^{-4}$  &$0.93\times10^{-4}$  &  &\\
Ref.~\cite{Gronau:2016xiq}  &$(1.94\pm0.70)\times10^{-3}~^\text{[a]}$  &$(3.86\pm1.35)\times10^{-3}~^\text{[a]}$  &$$  &$$\\
Ref.~\cite{Niu:2021qcc}  &$(5.8\pm2.1)\times10^{-3}$  &$(11.1\pm4.0)\times10^{-3}$  &  &\\
Ref.~\cite{Cheng:2022kea}  &$(1.76^{+0.18}_{-0.12})\times10^{-3}$  &$(3.03^{+0.29}_{-0.22})\times10^{-3}$  &$0.51\times10^{-3}$  &$0.28\times10^{-3}$\\
Ref.~\cite{Cheng:2022jbr}  &$(7.2\pm0.7)\times10^{-3}$  &$(13.8\pm1.4)\times10^{-3}$  &$(2.0\pm0.2)\times10^{-3}$  &$(1.1\pm0.1)\times10^{-3}$\\
Ref.~\cite{Ivanov:2023wir}  &$(5.4\pm1.1)\times10^{-3}$  &  &  &\\
Ref.~\cite{Faller:2015oma}  &$<3\times10^{-3}$  &$<6\times10^{-3}$  &$<3.7\times10^{-6}$  &$<1.1\times10^{-6}$\\
LHCb~\cite{LHCb:2020gge} &$(5.5\pm0.02\pm0.18)\times10^{-3}$  &  &  &\\
Belle~\cite{Belle:2022kqi} &$(5.4\pm0.5\pm0.5\pm1.2)\times10^{-3}$  &  &  &\\
\bottomrule[0.5pt]
\bottomrule[1pt]
\end{tabular*}
\begin{tablenotes}
\footnotesize
\item[a] The branching fractions were obtained by assuming constructive interference for amplitudes $A_{s\to u\bar{u}d}$ and $A_{cs\to cd}$.
\end{tablenotes}
\end{threeparttable}
\end{table*}

In addition, we present the predictions for the asymmetry parameters from various theoretical models in Table~\ref{tab:alpha}. Both the branching fractions and asymmetry parameters of the HFC weak decays $\Xi_{c}\to\Lambda_c\pi$ and $\Omega_{c}\to\Xi_c\pi$ are measurable at LHCb, Belle II, and BESIII experiments. We look forward to the possibility of testing these predictions in future experiments.

\begin{table*}[htbp]
\begin{center}
\caption{The comparison of different theoretical predictions for the asymmetry parameters $\alpha$ of the relevant HFC weak decays.}
\label{tab:alpha}
\renewcommand\arraystretch{1.25}
\begin{tabular*}{140mm}{c@{\extracolsep{\fill}}ccccc}
\toprule[1pt]
\toprule[0.5pt]
Asymmetry parameters  &$\Xi_{c}^{0}\to\Lambda_{c}^{+}\pi^{-}$  &$\Xi_{c}^{+}\to\Lambda_{c}^{+}\pi^{0}$ &$\Omega_{c}^{0}\to\Xi_{c}^{+}\pi^{-}$  &$\Omega_{c}^{0}\to\Xi_{c}^{0}\pi^{0}$\\
\midrule[0.5pt]
This work  &$-0.715\sim0.651$  &$-0.682\sim0.569$  &$-0.748\sim-0.532$  &$-0.457\sim-0.067$\\
Ref.~\cite{Cheng:2022kea}  &$0.70^{+0.13}_{-0.17}$  &$0.74^{+0.11}_{-0.16}$  &$-0.98$  &$-0.99$\\
Ref.~\cite{Cheng:2022jbr}  &$0.46\pm0.05$  &$0.45\pm0.05$  &$\approx-1.00$  &$\approx-1.00$\\
Ref.~\cite{Ivanov:2023wir}  &$-0.75$  &  &  &\\
\bottomrule[0.5pt]
\bottomrule[1pt]
\end{tabular*}
\end{center}
\end{table*}

\section{Summary}
\label{sec5}

The study of weak decays of charmed baryons provides valuable insights into both the weak decay mechanisms and nonperturbative effects in QCD. In this work, we investigate the HFC weak decays $\Xi_{c}\to\Lambda_c\pi$ and $\Omega_{c}\to\Xi_c\pi$, taking into account the $\Xi_{c}-\Xi_{c}^{\prime}$ mixing effect.

To calculate the decay amplitudes, we evaluate the factorizable contributions using the na\"{i}ve factorization approach, while the nonfactorizable contributions are estimated through the pole model and simplified using the soft-pion approximation. Additionally, the relevant nonperturbative parameters are determined within the framework of NRQM, where we directly use the exact baryon spatial wave functions derived from solving the Schr\"{o}dinger equation with a nonrelativistic potential, assisted by the Gaussian expansion method. This approach avoids the oversimplification of using Gaussian-type wave functions and reduces uncertainties associated with the choice of baryon wave functions.

With the computed nonperturbative parameters, we obtain the decay amplitudes and further explore the physical observables of these nonleptonic decays. Our results show that, for the mixing angle $\theta\in(24.4^\circ, 32.4^\circ)$, the experimental value $\mathcal{B}(\Xi_{c}^{0}\to\Lambda_{c}^{+}\pi^{-})=(0.55\pm0.11)\%$ can be well reproduced. We also estimate other branching fractions: $\mathcal{B}(\Xi_{c}^{+}\to\Lambda_{c}^{+}\pi^{0})=(8.69\sim9.79)\times10^{-3}$, $\mathcal{B}(\Omega_{c}^{0}\to\Xi_{c}^{+}\pi^{-})=(11.3\sim12.1)\times10^{-3}$, and $\mathcal{B}(\Omega_{c}^{0}\to\Xi_{c}^{0}\pi^{0})=(4.67\sim5.23)\times10^{-3}$. We expect that these predicted branching fractions can be tested in future experiments. Furthermore, we investigate the asymmetry parameters and find that our results are sensitive to the mixing angle. The measurement of these asymmetry parameters will be valuable for further constraining the mixing angle in the future.

\section*{ACKNOWLEDGMENTS}

Y.-S. Li is supported by the National Nature Science Foundation of China under Grant No. 12447155, and by the Postdoctoral Fellowship Program of CPSF under Grant No. GZC20240056. This work is also supported by the National Natural Science Foundation of China under Grant Nos. 12335001, and 12247101,  the ‘111 Center’ under Grant No. B20063, the Natural Science Foundation of Gansu Province (No. 22JR5RA389), the fundamental Research Funds for the Central Universities, and the project for top-notch innovative talents of Gansu province.

\end{document}